# XLMOD: Cross-linking and chromatography derivatization reagents ontology

**Authors:** Gerhard Mayer[1]

1 Ruhr University Bochum, Faculty of Medicine, Medizinisches Proteom-Center, D-44801 Bochum, Germany

**Abbreviations:**

| | |
|---|---|
| CID | Collision Induced Dissociation |
| CV | Controlled Vocabulary |
| CX-MS | Cross-Linking-Mass Spectrometry |
| Da | Dalton (atomic mass unit) |
| ESI | ElectroSpray Ionization |
| HUPO-PSI | HUman Proteome Organization – Proteomics Standards Initiative |
| IUPAC | International Union of Pure and Applied Chemistry |
| InChI | International Chemical Identifier |
| MS | Mass Spectrometry |
| NMR | Nuclear Magnetic Resonance |
| SMILES | Simplified Molecular Input Line Entry Specification |
| XRC | X-Ray Crystallography |

**Summary/Abstract:**

Mass spectrometry has experienced a rapid development since its first application for protein analysis in the 1980s. While the most common use of mass spectrometry for protein analysis is identification and quantification workflows on peptides (digested from their parent protein), there is also a rapidly growing use of mass spectrometry for "structural proteomics". One example is the analysis of cross-linked proteins that can give valuable structural information, complementing the information gained by classical protein structure determination methods, useful for integrated methods of structure determination and modeling. For a broad and reproducible application of cross-linking mass spectrometry a standardized representation of cross-linking experimental results is necessary. This paper describes the developing and release of the xlmod ontology from the HUPO-PSI. xlmod contains terms for the description of reagents used in cross-linking experiments and of cross-linker related chemical modifications together with their main properties relevant for planning and performing cross-linking experiments. We also describe how xlmod is used within the new release of HUPO-PSI's mzIdentML data standard, for reporting the used cross-linking reagents and results in a consistent manner. In addition xlmod contains terms for GC-MS and LC-MS derivatization reagents for specifying them in the upcoming mzTab-M and mzTab-L formats.

**Keywords:** Cross-linking reagents, cross-linker related chemical modifications, distance constraint, derivatization, xlmod ontology, mzIdentML 1.2, mzTab

**Download link:** https://raw.githubusercontent.com/HUPO-PSI/mzIdentML/master/cv/XLMOD.obo

## 1. Introduction

Traditionally proteins investigations are separated into two domains: At first, there are plenty of experimental methods for isolation, purification, identification and quantification of proteins (e.g. Difference Gel Electrophoresis (DIGE), SDS-PAGE (Sodium Dodecyl Sulfate – PolyacrylAmide Gel Electrophoresis), mass spectrometry (MS)). These aim at characterizing the proteins by their properties like e.g. their mass and sequence, their modifications, their physicochemical properties, interaction partners and so on. Second, there is the area of protein structural biology, aiming at the determination of tertiary and quaternary three-dimensional structure of the proteins, which is essential for their functional and dynamical behavior. With the technological progress of protein mass spectrometry (MS) in the last 3 decades MS is becoming more and more a useful tool for structural biology by giving information like dynamical, spatial and accessibility constraints, which can either be combined with other existing experimental protein structure determination or with computer modelling methods [1, 2]. Especially cross-linking mass spectrometry (CX-MS) [3, 4] is able to complement traditional methods like X-ray crystallography (XRC), nuclear magnetic resonance (NMR) and cryo-electron microscopy (cryo-EM), since each of these methods has some limitations: X-ray crystallography and NMR are both time-consuming and costly methods needing a relatively large amount of pure proteins. Furthermore they are either difficult to apply for membrane proteins and analyze the protein not in their native conformation (XRC), can be used only for smaller soluble proteins (NMR) or give only medium resolution information (cryo-EM), whereas CX-MS is a relatively fast high-throughput (HT) method by which also small amounts of proteins from protein mixtures can be analyzed.

The cross-linking reagents used in CX-MS contain reactive groups having specific reactivity's for reactions with the targeted functional groups of the amino acids and possessing spacer arms of varying lengths. Supplementary Table S5 shows some typical examples of such spacer arms. If such a cross-linking reagent reacts with residues from the same protein, one can derive distance constraints between these residues. These constraints are used as input for structural modeling. If the reagent links two different proteins, one can get information about the protein interaction partners and about their quaternary structure. Since the cross-links are covalent bonds, this allows also to detect transiently and weak binding interacting proteins. If the cross-linking reaction is terminated by a quencher, then the proteins contain also cross-linker specific, so-called dead end modifications. In general one distinguishes three types of cross-links [5]: Type 0 cross-links are cross-linker related dead end modifications, also called mono-links. They can give information about the solvent accessibility of the modified residues. Type 1 cross-links are intra-linked (intra-peptide or intra-protein) cross-links (also called loop-links). From intra-protein cross-links one can derive distance constraints for the maximum distance between the two cross-linked residues inside the protein. Type 2 cross-links are inter-linked (inter-peptide or inter-protein) cross-links. Inter-protein cross-links can give information about interacting protein partners and deliver distance constraints between subunits of protein complexes.

Since the efficiency of the cross-linking reactions is low, the cross-linked peptides occur in low abundance. Therefore, an enrichment step after the enzymatic digestion, e.g. a size exclusion chromatography (SCX), is done before the cross-linked peptides are identified [6]. Another possibility is to use trifunctional cross-linkers, which contain a handle, e.g. biotin, which can later be used to separate the cross-linked peptides by affinity chromatography [7].

The identification of the cross-linked peptides by a search engine is challenging, because the identification of the cross-links has in general quadratic computational complexity compared with the linear complexity of a database search in normal MS [8]. This is due to the fact that one has to search for all possible binary combinations of peptides. This can either be done in an exhaustive search [9] or one explicitly encodes all the cross-linked peptides in the search database [10]. Therefore cross-linkers with certain properties, which can be utilized to develop specialized and faster search strategies, were developed. Examples are CID-cleavable cross-linkers [11, 12], isotope-labelled cross-linkers [13, 14], Protein Interaction Reporters (PIR) [15], photo-cleavable cross-linkers or cross-linker with MS2 labile bonds, allowing to identify the cross-linked peptides in the following MS3 step [16]. Another category is photoactivatable cross-linkers [17-19], which are mostly not selective regarding the targeted amino acid, but allow controlling the cross-linking reaction using a UV light source. Aldehydes such as glutardialdehyde or formaldehyde are broadly specific, i.e. they react with many amino acid residues and also with DNA and can also be used for the detection of protein-DNA interactions [20].

## 2. Cross-linking reagents

Besides deriving distance constraints for structure determination, cross-linking is useful to study receptor-ligand interactions, for the location determination of membrane proteins and cell surface receptors, for the immobilization on solid surfaces, for the production of immunotoxins and for labelling of proteins with fluorescent dyes (Table 1). For instance, the photo-affinity labeling method gets three-dimensional information about the binding interfaces, either between two proteins or between a protein and a drug in the binding pocket of that protein receptor for a small molecule drug [21]. This method uses cross-linkers with the photo-reactive groups (azido, benzophenone and diazirines) mentioned above. These cross-linkers are able to react with C-H bonds, which is the reason for the broad non-selective reactivity of these photoactivatable cross-linkers [21, 22]. In addition, there are cross-linkers available, which can link proteins with nucleic acids, lipids or carbohydrates. For instance cross-linkers with the photo-reactive psoralen group react with pyrimidines, especially the thymines of nucleic acids [23] and diazirine reactive groups are able to react with aldehydes of sugar chains [24].

Another possibility is the site-specific incorporation of cross-linkers into a cell's proteome by using photo-reactive amino acids analogues in order to study protein-protein interactions (PPIs) in living cells (*in vivo* crosslinking) [25]. One possibility is to genetically encode - by using an expanded genetic code - for the incorporation of a photo-reactive amino acid analogue like e.g. p-benzoyl-L-phenylalanine (pBpa) into the targeted protein [26-28]. For easy identification by doublets at M and M + 11, a deuterium-labelled pBpa-d11 version is used [29].

Other photo-reactive amino acids analogues that can be used are photo-L-Leu, photo-L-Ile and photo-L-Met [30] containing a photo-reactive diazirine group by which they can insert into C-H, N-H or S-H bonds, where no structural changes of the peptides due to the insertion of the diazirine ring were observed [31]. For all of them it was shown that their incorporation retains the protein structure and functions [28].

Table 1: Example use cases for cross-linking reagents usage.

| Application | Comment |
| --- | --- |
| Deriving distance constraints for investigation of protein structure and guiding a structure-based modelling process. | Uses cross-linkers with different chemical reactivity and with different spacer lengths to derive a set of distance constraints which can restrict the search space in a subsequent structure modelling step. Often isotopically labelled heavy/light cross-linker pairs are used, since then the cross-linked peptides can be easily identified by the search software due to their characteristic doublet signals in the MS spectrum. Low concentration of protein and high cross-linker concentration favors intramolecular crosslinking. For the identification of subunits, cleavable cross-linkers are useful, so that by 2D-gelelectrphoresis the subunits of the complex can be detected. |
| Study of protein-protein or protein-DNA and receptor-ligand interactions for the investigation of signal transduction. | Uses protein cross-linkers to establish an association between two interacting proteins resp. a receptor protein and its ligand for the derivation of the quaternary structure of protein complexes [32]. |
| Determination of the location of membrane proteins and identification of cell surface receptors. | Integral membrane proteins react with water-insoluble cross-linkers, but not with hydrophilic cross-linkers. Membrane impermeable cross-linkers like Sulfo-NHS esters and pegylated cross-linkers, due to their hydrophilic nature, react only with proteins located on the cell surface. |
| Immobilization of proteins on solid supports. | Typical supports are e.g. nitrocellulose membranes, glass, agarose, polystyrene plates or beads. Allows e.g. to produce protein nanosurfaces for the production of biosensors or analytical devices like e.g. protein chips [33]. |
| Use as immunological tools, e.g. for the production of immunotoxins or the coupling of haptens to carrier proteins for getting antigens for antibody production. | Linking toxins to antibodies for getting molecules that can be used for targeted killing of cancer cells. Often cross-linkers with a cleavable disulfide S-S bond are used, because most cells are able to cleave these and thereby the bound toxin is released. |
| Transfer of labels or reactive groups | Cross-linkers can also be used to modify target groups on a |

| | protein or to add spacers for later coupling reactions. |
|---|---|
| Labeling of proteins. | Enable identification, enrichment, purification or detection of proteins by transferring labels like e.g. biotin or fluorescent dyes from a known bait protein to an unknown interacting prey protein. |
| *In vivo* crosslinking [34]. | Stabilizing transient endogenous protein-protein or receptor-ligand interactions. Photo-reactive amino acids analogues are used to incorporate the cross-linkers into the cell's proteome. |

## 3. Properties and selection criteria for choosing cross-linking reagents

Criteria for the selection of a cross-linker were defined by Jin Lee [35] and Back et.al. [36]. The most important criteria are the specificity, which is given by the reactive groups, and the spacer lengths, especially with respect to getting distance constraints for later protein structure modeling.

Further possible criteria, e.g. the ability to permeate membranes, determine which kinds of proteins can be cross-linked. For an easier detectability of the cross-linked peptides by the used search machines, the availability of isotopically coded versions of the cross-linker or the cleavability of the cross-linker can be important, since then such cross-linked peptides are more easily detectable by the used search engines. When such cross-linkers are not available, one can make use of trifunctional cross-linkers with an additional handle, which allow for an additional enrichment step before the mass spectrometry analysis. Another possibility is the enrichment by using size exclusion chromatography [37]. Table 2 gives an overview about cross-linker properties, documented in the xlmod ontology and useful for cross-linker selection. In order to expand the applicability of the xlmod ontology, we included also some properties, which are not specific for XL-MS. Some vendors also offer cross-linker selection guides and/or interactive selection tools for choosing proper cross-linkers (see Supplementary Table S3).

Table 2: Cross-linker properties

| **Property** | **Comment / Reason / Example** |
|---|---|
| Chemical specificity | The specificity is determined by the reactive groups, targeting specific groups of the different amino acids. |
| Reaction conditions | e.g. buffer system, pH, concentrations, … |
| Spacer length | Distance between the cross-linked proteins acting as molecular rulers for deriving spatial constraints for protein structure modelling [38]. |
| Spacer composition | Determines hydrophobicity. For instance, a spacer containing PEG (polyethylenglycol) makes the cross-linker hydrophobic and therefore prevents hydrophobic aggregation. |

| Spacer cleavability | If the cross-linker can be cleaved, either chemically, photo-activated or by fragmentation, e.g. CID (Collision Induced Dissociation). This cleavability e.g. allows the release of labeled cross-linked peptides from the supports used in an affinity based purification step [39]. |
|---|---|
| Water solubility / membrane permeability | Determines if the cross-linker can detect cell surface proteins or can cross-link integral membrane proteins (which are hydrophobic) resp. if a cross-linker can permeate into cells (e.g. for in vivo cross-linking). |
| Isotope-coded versions | Makes the cross-link identification process by the search program easier, e.g. by searching for the characteristic doublets [29] of cross-links containing the heavy resp. light version of the cross-linker. |
| Selectivity of the reactive group | If spontaneously reactive with the targeted functional groups of residues or photo-reactive (non-selective), e.g. for two-step cross-linking reactions. |
| Handle for detection / identification or purification / enrichment | Using cross-linkers with a handle (e.g. biotin or a fluorophore) for easy enrichment of the cross-linked peptides. |

## 4. Derivatization reagents

Because derivatization reagents like cross-linker have functional groups specifically reacting with chemical target groups, their description is very similar to that of the cross-linker reagents. Therefore, we decided to include them into the XLMOD ontology, so that beside the cross-linker reagents used in proteomics we also included derivatization reagents, useful for annotation in metabolomics and lipidomics. These derivatization reagents are either used in GC/MS for enhancing the volatility of polar metabolites, allowing them to pass easily into the gas phase [40, 41]. Similarly in LC-MS approaches the derivatization is done in order to enhance the LC signal intensities, the ionization in the ESI, (ElectroSpray Ionization) and/or the MS-detectability of the metabolite by a more efficient fragmentation [42]. The used derivatization reagents should specifically react with the targeted metabolites producing a high yield of the derivative [42]. The main groups of derivatization reagents for GC-MS are acylation, alkylation and silylation reagents, which replace active hydrogens (e.g. from -SH, -OH, -NH, -COOH groups) by an acyl, alkyl resp. silyl group, which confers the derivatized target molecule a more hydrophobic character, which again enhances their volatility.

Derivatization reagents for LC-MS at one hand have a functional group for the formation of the derivative, and on the other hand they introduce a group, which allows an easier detection of the derivative. This improved detectability is accomplished either in the LC part by introducing a chromophore, a fluorophore or an electrophore group for detection by UV/VIS light, by fluorescence or by electrochemical detectors [43]. Another possibility is to enhance the detectability in the MS part by improved ion formation in the ESI or by making the fragmentation more efficient. In mzTab-M [44] the used derivatization reagents should be reported in the metadata section under derivatization_agent[1-n] as specified by the mzTab-M specification document available at https://github.com/HUPO-PSI/mzTab.

## 5. xlmod ontology for cross-linking and derivatization reagents

Although there are several articles listing cross-linkers for use in structural proteomics [45-47], there was still an urgent need for a structured ontology, which can serve both as a database for choosing proper cross-linkers, as well as source for uniquely describing identified cross-linked peptides and peptides with post-translational modifications derived from cross-linkers.

For building up the xlmod ontology we scanned vendor websites, publications and catalogues (see Supplementary Tables S1 - S3), relevant papers from the PubMed literature database [48] and some relevant books [49, 50].

The most important feature for the selection of a cross-linker reagent is its specificity for amino acids [51] and then it's spacer length. The amino acid residues, which are currently targetable by cross-linkers, are summarized in Supplementary Table S4.

It was shown, that the availability of cross-linking reagents targeting the acidic carboxyl groups of Asp and Glu allows deriving distance constraints that orthogonally complement constraints obtained from Lys cross-linking [51, 52]. It can be expected that cross-linkers specifically targeting other amino acids, e.g. the indole amine of tryptophan, can further enhance the power of the protein cross-linking method as a 3D structure determination tool.

The reactive groups occurring in cross-linkers and their amino acid specificities are reviewed in [25, 46, 47, 53], and summarized in Table 3.

Table 3: Overview about the reactive groups of cross-linking reagents

| Reactive group | Target functional group | Specificity for amino acids | Photo-reactive | Reference (PMID) |
|---|---|---|---|---|
| formaldehyde | primarily amines, but broadly specific | K,N,R,Q,Y, N-term secondary: C,F,H,S,T,W | | 25979347 |
| 1-hydroxy-7-azabenzotriazole | amines | K,N,Q,R,Protein N-term | | 19994840 |
| amine | carboxyl groups | D,E | | |
| aryl halide | primary amines | K,N,Q,R,Protein N-term | | |
| aryl azides (e.g. phenyl-, nitrophenyl-, | insertion into C-H and N-H bonds; | nonselective | yes | 22641729 |
| hydroxyphenyl-, tetrafluorophenyl,-perfluoroaryl azide) | primary amines | | | |
| azido-methylcoumarin | --- | nonselective | yes | |
| benzophenone | --- | nonselective | yes | |
| carbodiimide | carboxyl-to-amine reactive | (K,N,Q,R,Protein N-term)&(E,D,Protein C-term) | | |
| carbonyl | hydrazines | --- | | |
| carboxyl | amines | K,N,Q,R,Protein N-term | | |
| diazirine | --- | nonselective | yes | 21778064 |
| diazo group | --- | nonselective | yes | |
| dimethylether | carboxyls | D,E | | |
| glyoxal | arginine-specific | R | | |
| haloacetyl (chloro-, bromo-, iodoacetyl) | sulfhydryls | C | | |
| hydrazide | carbonyls | D,E,Protein C-term | | |
| hydrazone | carbonyls | D,E,Protein C-term | | |
| hydroxybenzotriazole | amines | K,N,Q,R,Protein N-term | | 19994840 |
| hydroxymethyl phosphine | amines | K,N,Q,R,Protein N-term | | |
| imidoester | amines | K,N,Q,R,Protein N-term | | |
| isocyanate | hydroxyl | S,T,Y | | |
| maleimide | sulfhydryl | | | |
| methanethiosulfonate | sulfhydryl | | | |
| N-hydroxyphthalimide | amines | K,N,Q,R,Protein N-term | | 19994840 |
| NHS (N-hydroxysuccinimide) ester | lysines and N-termini, but also hydroxyls | K,Protein N-term secondary: S,T,Y | | |
| pentafluorophenyl | amines | K,N,Q,R,Protein N-term | | 22067100 |
| platinum (II) | thioether, sulfhydryl, and | C,H,M | | 21591778 |

| | imidazole | | | |
|---|---|---|---|---|
| psoralen | pyrimidines, esp. thymins | Thy | | |
| pyridinyldisulfide (pyridyldithiol) | sulfhydryls | C | | |
| S-acetyl | incorporates sulfhydryls | --- | | 9177838 |
| Sulfo NHS (N-hydroxysulfosuccinimide) ester | lysines and N-termini, but also hydroxyls | K,Protein N-term secondary: S,T,Y | | |
| thioimidate | amines | K,N,Q,R,Protein N-term | | 20795639 |
| vinyl sulfone | amines, sulfhydryls, hydroxyls | C,K,N,Q,R,S,T,Y | | |

It must be mentioned here that the "specificities" for amino acids are not absolute, but rather describe a tendency for preferred reaction targets, because the specificities can vary with the reaction conditions [54]. For instance NHS esters and Sulfo NHS esters react primarily with the amine group of lysines and with protein N-terminals, but to a lower degree they can also react with the hydroxyl groups of serine, threonine and tyrosine [25, 55], as specified by the property "secondarySpecificities".

In general, each cross-linking reagent term has a unique and fixed id, a short name, a definition, an optional list of synonyms and a list of properties with their values, the specification of its parent terms and then a list of relationships, describing e.g. the chemical reactivity.

As a concrete example, the entry for the reactive group "NHS ester" is defined as follows:

> *[Term]*
> *id: xlmod:00101*
> *name: NHS ester*
> *def: "A reactive group (N-hydroxysuccinimide) that reacts with lysines and N-termini but also serines, threonines and tyrosines." [PSI:XL]*
> *synonym: "N-hydroxysuccinimide ester" EXACT []*
> *property_value: specificities: "(K,Protein N-term)" xsd:string*
> *property_value: secondarySpecificities: "(S,T,Y)" xsd:string*
> *is_a: xlmod:00100 ! reactive group*
> *relationship: has_property xlmod:00013 ! membrane permeable*
> *relationship: is_reactive_with xlmod:00028 ! primary amine reactive*

## 6. General structure of the xlmod ontology

The xlmod ontology consists of four main branches of cross-linker or derivatization reagent terms and defines several relations, which can be used to describe such terms in general (see Figure 1). The three main branches are:

- cross-linker related chemical modification
- cross-linking entity
- derivatization entity
- label transfer reagent

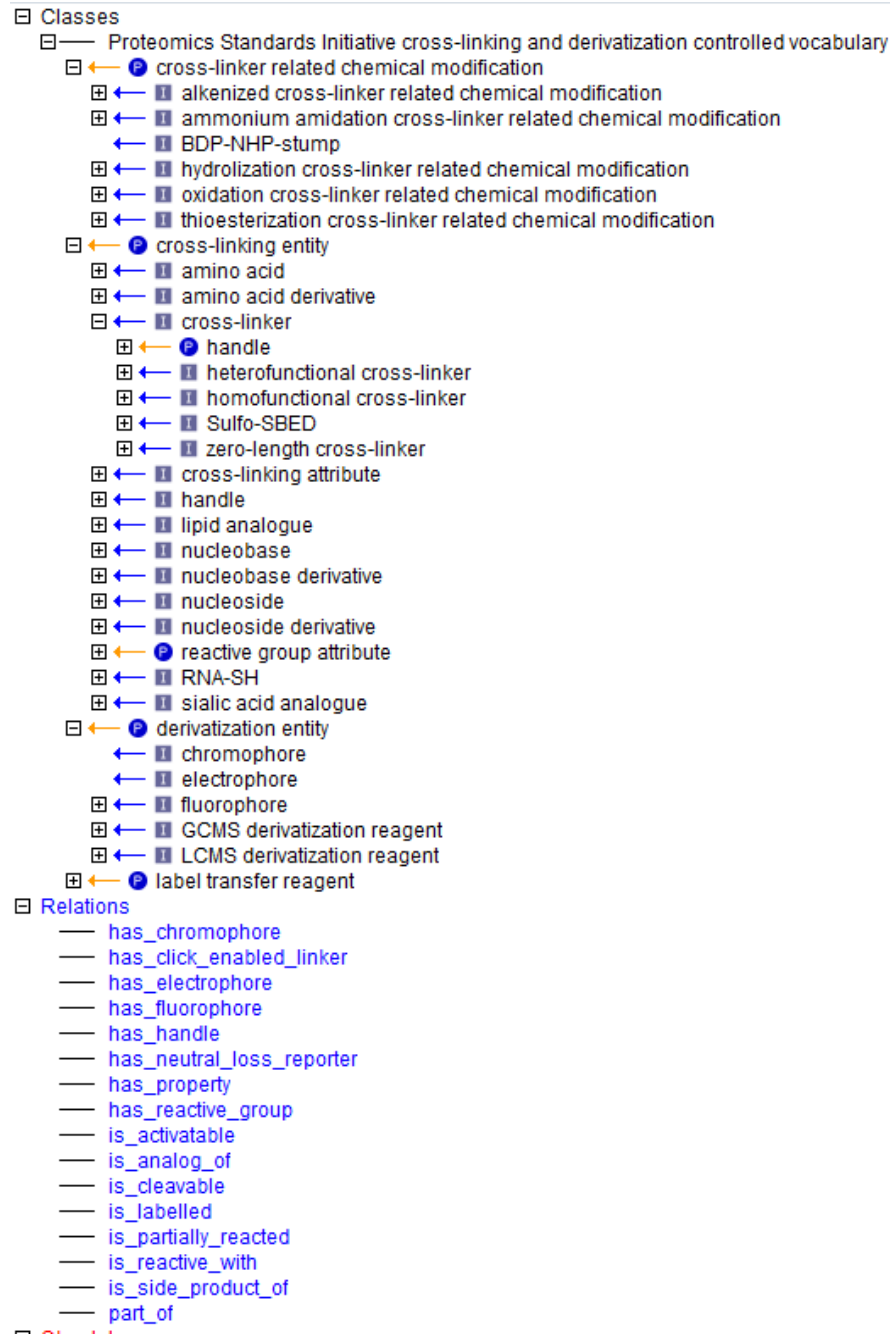

The *cross-linker related chemical modification* branch contains partially reacted products of cross-linkers, e.g. modifications resulting from dead-ends (mono-links), which are no real cross-links, since they do not link two chains of proteins. In principle one can quench the cross-linking reaction with any amine, but the mostly ammonium salts or Tris are used. Therefore mainly amidated or hydrolyzed cross-linkers, but also thioesterized or alkenized forms are listed in xlmod. The amidated modifications are subdivided into ammonium amidated and Tris amidated forms, since these differ slightly in the resulting modifications and their masses, because ammonium salts convert to -C(=O)-NH2, whereas the use of Tris(hydroxymethyl)-aminomethane as a quenching substance leads to the formation of dead ends of the form -C(=O)-NHC(CH2OH)3. At the moment we have no instances of Tris amidated forms, but we distinguished these cases to allow them to be added later. The label transfer reagents are either trifunctional cross-linkers containing a handle, e.g. biotin, for enrichment or for transferring a label to a prey protein.

The main branches under *cross-linking entity* are:
- amino acid and their derivatives (containing photo-reactive amino acid analogs)
- the cross-linkers in the narrower sense (resp. the cross-linked products derived from them)
- cross-linking attributes for a more detailed characterization of the cross-linkers, e.g. their cleavability, and cross-linker derived chemical modifications
- handles used for detection, affinity enrichment or purification of cross-linked peptides (see Supplementary Table S6)
- reactive group attributes for a more detailed description of reactive groups

Important cross-linking attributes are the reactive groups, which have specific chemical reactivity towards functional target groups, and which are listed under the reactive group attributes.

The branch *derivatization entity* contains sub-branches for
- chromophores (for UV/VIS detection)
- fluorophores (for fluorescence detection)
- electrophores (for electrochemical detection)
- GCMS derivatization reagents
- LCMS derivatization reagents.

A typical example for a cross-linker term is:

> *[Term]*
> *id: xlmod:02001*
> *name: DSS*
> *def: "Disuccinimidyl suberate." [CAS:68528-80-3, PubChem_Compound:100658, PMID:16944939, ChemSpiderID:90944, ChemicalBookNo:CB1219157]*
> *synonym: "Bis(succinimidyl) suberate" EXACT []*
> *synonym: "Suberic acid bis(N-hydroxysuccinimide ester)" EXACT []*
> *synonym: "Disuccinimidyl octanedioate" EXACT []*

*synonym: "DSS-d0" EXACT []*

*synonym: "DSS-H12" EXACT []*

*synonym: "1,1'-[(1,8-Dioxooctane-1,8-diyl)bis(oxy)]dipyrrolidine-2,5-dione" EXACT []*

*property_value: bridgeFormula: "C8 H10 O2" xsd:string*

*property_value: monoIsotopicMass: "138.06807961" xsd:double*

*property_value: reactionSites: "2" xsd:nonNegativeInteger*

*property_value: spacerLength: "11.4" xsd:float*

*is_a: xlmod:00005 ! homofunctional cross-linker*

*relationship: has_reactive_group xlmod:00101 ! NHS ester*

The cross-linker entries possess a unique ID and (short) name. Then follows a definition line containing the long name as well as a list of database cross references, which can be of one of the types listed in Table 4. The PMID cross references and web links can occur several times for a given definition, whereas the other possible cross references must be unique.

Table 4: Used database cross references (dbxref's) in the xlmod.obo ontology

| DB cross reference | DB name | Link |
| --- | --- | --- |
| Beilstein | Beilstein Registry Number | https://www.elsevier.com/solutions/reaxys |
| CAS | Chemical Abstract Service | https://www.cas.org/content/chemical-substances |
| CBNumber | Chemical Book | http://www.chemicalbook.com/Search_EN.aspx |
| ChemSpiderID | ChemSpider | http://www.chemspider.com |
| MDL | former Molecular Design Ltd., now BIOVIA | http://accelrys.com <br> https://www.3ds.com/products-services/biovia/ |
| PMID | Pubmed ID | http://www.ncbi.nlm.nih.gov/pubmed/advanced |
| PubChem_Compound | PubChem | https://www.ncbi.nlm.nih.gov/pccompound/advanced |
| http://... | Web link | specific web link, where the cross-linker is defined |
| DOI:… | Web link | web link, e.g. for publications not indexed in PubMed |

The chemical substance registry references can be used, if one wants to get more detailed information about a cross-linker, e.g. the IUPAC name, a full list of synonyms, the SMILES string, InChI key, 3D formula, and so on. From the Chemical Book repository, one can get also MDL .mol files containing the exact coordinates of the 3D structure. In principle one can then calculate the spacer arm lengths from the coordinates in these mol files [56], if one is acquainted with the reaction mechanism, so that one exactly knows which atoms belong to the spacer chain of the resulting cross-linked peptides. Details of that format are described at http://accelrys.com/products/collaborative-science/biovia-draw/ctfile-no-fee.html (accessed 08/2018) and http://media.accelrys.com/downloads/ctfile-formats/ctfile-formats.zip. Other possibilities to get the spacer arm length are to calculate them from force filed simulations [38] or to calculate them from 3D coordinates. If the coordinates are unknown, one can calculate them from the SMILES strings, which one can get from PubChem, by using cheminformatics programs like e.g. the 3D structure generator CORINA, available from http://www2.chemie.uni-erlangen.de/services/telespec/corina/ (accessed 08/2018) [57], which converts the SMILES string into 3D coordinates. Another possibility is to measure the spacer arm lengths with chemical drawing programs like e.g. BIOVIA Draw (http://accelrys.com/resource-center/downloads/freeware/, accessed 08/2018). A tool that can be used to calculate the m/z ratio of cross-linked peptides is the computer program GPMAW [58] (http://www.gpmaw.com/html/cross-linking.html and http://www.gpmaw.com/html/new_cross-linker.html, accessed 08/2018).

Further available software tools for cross-linking analysis and identification are already summarized by Mayne and Patterton [59], Sinz [60], Leitner *et.al.* [16], Tran *et.al.* [32], and in the list of cross-linking software tools at https://omictools.com/cross-linked-peptide-identification-category (accessed 08/2018).

In the xlmod ontology we only listed synonyms, which are commonly used by the cross-linking community. The bridge formula (sum formula of the spacer arm) is given in Hill notation [61]. In addition, the monoisotopic mass calculated from the bridge formula, is specified. Then the number of reaction sides is given and the cross-linker is either classified as homofunctional (i.e. the cross-linker has only one sort of reactive group) or heterofunctional (i.e. containing more than one different reactive groups). In the DSS example above the number of reaction sites is 2, so that it's a homobifunctional cross-linker. At the end the type of reactive groups are listed.

In order to avoid redundancy, the specificities for the amino acids are not repeated in the term for the cross-linker, so that the specificities must be looked up at the terms for the listed reactive groups of that cross-linker. This means that the properties from reactive groups are inherited and apply to the cross-linking reagents containing the respective reactive group. For example, it follows that a cross-linker with a NHS ester reactive tends to be more membrane permeable, whereas one with a Sulfo-NHS ester reactive group is more hydrophilic. The overall hydrophobicity is also influenced by the hydrophilicity of the spacers and affinity arms, e.g. PEG spacers make a reagent the more hydrophilic the longer the PEG spacer group is. Likewise, it is clear, that a cross-linker with a pyridinyldisulfide reactive group possesses a cleavable S-S bond or that a cross-linker with a biotin handle is enrichable.

An example for a cross-linker related post-translational modification is the term for "hydrolyzed DSS", where the relationship "is_side-product_of" specifies from what cross-linker reagent the chemical modification is derived from:

> *[Term]*
> *id: xlmod:01002*
> *name: hydrolyzed DSS*

>    def: "Hydrolyzed disuccinimidyl suberate." [PSI:XL]
>    property_value: deadEndFormula: "C8 H12 O3" xsd:string
>    property_value: monoIsotopicMass: "156.07864431" xsd:double
>    is_a: xlmod:00096 ! hydrolization cross-linker related chemical modification
>    relationship: is_partially_reacted xlmod:00011 ! hydrolyzed
>    relationship: is_side_product_of xlmod:02001 ! DSS

The following is an example for a labelled cross-linker with the relationship "is_labelled":

>    [Term]
>    id: xlmod:02002
>    name: DSS-d4
>    def: "Deuterium labelled disuccinimidyl 2,2,7,7-suberate." [PubChem_Compound:91757798, PMID:11354472]
>    property_value: bridgeFormula: "C8 D4 H6 O2" xsd:string
>    property_value: doubletDeltaMass: "4.02508" xsd:double
>    property_value: monoIsotopicMass: "142.093186586" xsd:double
>    property_value: reactionSites: "2" xsd:nonNegativeInteger
>    property_value: spacerLength: "11.4" xsd:float
>    is_a: xlmod:00005 ! homofunctional cross-linker
>    relationship: has_reactive_group xlmod:00101 ! NHS ester
>    relationship: is_labelled xlmod:00010 ! deuterium labelled

The property "doubletDeltaMass" gives the m/z difference between the labelled and the unlabeled versions of the cross-linker and can be used by cross-linking search-engines for easy detection of the corresponding cross-linked peptides.

As an example for an entry of a photo-reactive amino acid analogue the entry for pBpa-d11 is specified as:

>    [Term]
>    id: xlmod:01904
>    name: pBpa-d11
>    def: "Deuterium labelled p-benzoyl-L-phenylalanine." [PMID:18704231]
>    comment: A photoreactive deuterium-labelled amino acid analog of L-Phenylalanine for incorporation during protein synthesis that can be used for in vivo labelling, cross-linking and protein-protein interaction studies in live cells.
>    property_value: reactionSites: "1" xsd:nonNegativeInteger
>    is_a: xlmod:00049 ! amino acid derivative

> *relationship: has_reactive_group xlmod:00119 ! benzophenone*
> *relationship: is_activatable xlmod:00149 ! photoactivatable*
> *relationship: is_analog_of xlmod:00070 ! L-Phenylalanine*
> *relationship: is_labelled xlmod:00010 ! deuterium labelled*

Since the PSI-MOD ontology is not maintained anymore, we also added a special branch to the xlmod ontology that contains label transfer reagents, which are primarily used for transferring labels and not for cross-linking. An example would be

> *[Term]*
> *id: xlmod:01800*
> *name: Psoralen-PEG3-Biotin*
> *def: "Psoralen-PEG3-Biotin biotinylation reagent." [PSI:XL]*
> *property_value: hydrophilicPEGchain: "3" xsd:nonNegativeInteger*
> *property_value: reactionSites: "1" xsd:nonNegativeInteger*
> *property_value: spacerLength: "36.86" xsd:float*
> *is_a: xlmod:00003 ! label transfer reagent*
> *relationship: has_handle xlmod:00051 ! biotin*
> *relationship: has_property xlmod:00014 ! hydrophilic*
> *relationship: has_reactive_group xlmod:00134 ! psoralen*

At the moment the xlmod ontology contains 16 type definitions and 1083 terms, of which 248 are bifunctional cross-linker terms, 103 terms belong to the class of cross-linker related chemical modifications and 84 terms describe chemical reactivities, reactive groups and reactive group attributes. In addition there are 174 terms for GC/MS derivatization reagents and 324 terms for LC/MS derivatization reagents. The other terms describe zero-length and trifunctional cross-linkers, label transfer reagents, amino acids and their derivatives, handles, chemical reactivities and other attributes and properties of the cross-linkers. For details see the Supplementary Table S7 and S8.

## 7. Reporting of cross-linking results

Controlled vocabulary (CV) terms from the xlmod ontology and from the psi-ms ontology [62] are used together for documenting cross-linking searches in version 1.2 of the XML-based mzIdentML standard format [63], see Figure 2. For that one has to specify the CV term "crosslinking search" (MS:1002494) in the <SpectrumIdentificationProtocol> element of the mzIdentML file [63]. The two cross-linked peptides must have <Modification> elements, which are marked as cross-linked peptides by using the CV terms "cross-link donor" and "cross-link acceptor", which are defined in the psi-ms ontology [62]. These two corresponding peptides are matched together by having the same value. The mzIdentML validator [64] checks that these two CV terms, grouped together by the same value, are located under different <Modification> elements. By convention, the donor peptide is the one with the longer peptide sequence. In case that both peptides have the same length, the donor is defined as the one, which comes first in the alphabetical order of the peptide sequence.

```xml
<Peptide id="PEP_1">
    <PeptideSequence>STMLEKIK</PeptideSequence>
    ...
    <Modification location="6" residues="K" monoisotopicMassDelta="138.06807961">
        <cvParam accession="xlmod:02001" cvRef="PSI-MS" name="DSS"/>
        <cvParam accession="MS:1002509" cvRef="PSI-MS" name="cross-link donor" value="CLM_1"/>
    </Modification>
</Peptide>

<Peptide id="PEP_2">
    <PeptideSequence>LCHKNK</PeptideSequence>
    ...
    <Modification location="5" residues="K" monoisotopicMassDelta="0">
        <cvParam accession="MS:1002510" cvRef="PSI-MS" name="cross-link acceptor" value="CLM_1"/>
    </Modification>
</Peptide>

<SpectrumIdentificationProtocol id="SIP_1" analysisSoftware_ref="AS_1">
    ...
    <AdditionalSearchParams>
        <cvParam accession="MS:1002494" cvRef="PSI-MS" name="cross-linking search"/>
    ...
</SpectrumIdentificationProtocol>

<SpectrumIdentificationResult spectraData_ref="SDATA_1" spectrumID="scan=1,scan=2" id="SIR_1">
    <SpectrumIdentificationItem passThreshold="1" rank="1" peptide_ref="PEP_1" experimentalMassToCharge="672.374450" chargeState="3" id="SII_1">
        ...
        <cvParam accession="MS:1002511" cvRef="PSI-MS" name="cross-link spectrum identification item" value="CL_SII_1"/>
        <cvParam accession="MS:1002545" cvRef="PSI-MS" name="xi:score" value="1.23"/>
    </SpectrumIdentificationItem>

    <SpectrumIdentificationItem passThreshold="1" rank="1" peptide_ref="PEP_2" experimentalMassToCharge="676.400268" chargeState="3" id="SII_2">
        ...
        <cvParam accession="MS:1002511" cvRef="PSI-MS" name="cross-link spectrum identification item" value="CL_SII_1"/>
        <cvParam accession="MS:1002545" cvRef="PSI-MS" name="xi:score" value="1.23"/>
    </SpectrumIdentificationItem>
</SpectrumIdentificationResult>
```

XL Donor specifies the cross-linking reagent

XL Acceptor reported with monoisotopicMassDelta of 0

By convention, the cross-link acceptors are reported by a <Modification> element with a monoisotopic mass delta of zero, and the cross-link donor with the actual mass of the cross-link. In addition the peptide, which is the "cross-link donor" must have another CV term annotating its <Modification> element [63]. This term must be a term located under the branch "cross-linker" (xlmod:00004) from the xlmod ontology, and describes the cross-linking reagent. In contrast, a dead-end modification is annotated by only one CV term, referencing a child term of "cross-linker related chemical modification" (xlmod:00002) from the xlmod ontology.

Since there are always pairs of donor and receiver peptides, there must be two <PeptideEvidenceRef> elements (referencing the cross-linked peptides), that occur paired with the same charge state in two <SpectrumIdentificationItem> elements under the same <SpectrumIdentificationResult> element with the same value for the CV term "cross-link spectrum identification item" (MS:1002511) annotating these <SpectrumIdentificationItem> elements. If one uses isotope-labelled cross-linker reagents, there are even four <SpectrumIdentificationItem> elements for the same cross-link and charge state, since donor and acceptor can occur now cross-linked with the heavy and light version of the used cross-linker reagent. This rule is also checked by the mzIdentML validator [64].

We want to mention here, that beside mzIdentML 1.2 there is also an alternative open data format for reporting cross-linking results, developed by Hoopmann et.al. [65] and is an extension of the pepXML format [66].

## 8. Outlook to the further development of xlmod

Then HUPO-PSI intends to steadily further develop the xlmod ontology. Currently, on the one hand there are still some cross-linkers missing and on the other hand for some cross-linkers there are still some data missing. Our hope is that the cross-linking community will contribute such missing data and request terms for still missing respective newly developed cross-linkers via the 'psidev-ms-vocab' mailing list at https://lists.sourceforge.net/lists/listinfo/psidev-ms-vocab and thereby contribute to make this ontology a standard source for cross-linkers, their related chemical modifications, and derivatization reagents. Then this ontology can be utilized by specialized cross-linking identification and/or analysis software like e.g. xiFDR [67], SIM-XL [68] or OpenMS [69], which allow to report their results in the mzIdentML 1.2 [63] standard format. The xlmod ontology is also open for bio-conjugative reagents used in cross-linking of either RNA-RNA [70], or proteins to other biomolecules like e.g. DNA / RNA [23, 71-73], lipids [74, 75] or sugars [76-78]. For instance there are lipid analogues defined, which are useful for describing the interaction of lipids with membrane proteins [75], the terms for nucleobases and their derivatives can be used for the description of protein-RNA interactions [72] and the sialic acid analogue terms are intended for the description of glycan-protein crosslinking [78] experiments.

Conceivable is the use of xlmod as a database for the development of an open-source tool for the planning of cross-linking experiments, allowing the selection of the most promising cross-linking reagents dependent of the detailed aims of the planned experiment. For the easy access to xlmod the development of a general Java access library, resolving all the inheritance dependencies between the different terms, e.g. between the reactive groups and the cross-linking reagents, would also be desirable.

Desirable for the reproducible application of CX-MS in integrated structure determination and prediction approaches are also standardized experimental protocols, since the specificities of the cross-linking reagents depends on the reaction conditions [79], as well as robust, standardized and automatable data analysis workflows [80, 81]for such integrated structure determination and modeling methods. The xlmod ontology is one building block for such a future highly automated high-throughput application of CX-MS in structural biology.

**Acknowledgements:**

Gerhard Mayer was funded by the BMBF grant de.NBI - German Network for Bioinformatics Infrastructure (FKZ 031 A 534A).

Supplementary Table S1: Used vendor web page links.

| Vendor | Web site (accessed 02/2020) |
|---|---|
| Santa Cruz Biotechnology (SCB) | http://www.scbt.com/chemicals-table-crosslinkers.html |
| SCB | http://www.scbio.de/chemicals-table-crosslinkers.html |
| ChemCruz | http://www.scbio.de/chemicals-table-homobifunctional_crosslinkers.html |
| Invitrogen | https://www.thermofisher.com/search/global/searchAction.action?query=crosslinkers |
| Sigma Aldrich | http://www.sigmaaldrich.com/catalog/search?interface=All&term=crosslinker&N=0&page=2&mode=mode+matchpartialmax&focus=product |
| Sigma Aldrich | http://www.sigmaaldrich.com/life-science/molecular-biology/molecular-biology-products.html?TablePage=9621740 |
| Clear Synth | http://www.clearsynth.ca/search.asp?category=Cross%20Linking%20Reagents |
| ThermoFisher | https://www.thermofisher.com/search/results?query=crosslinkers&resultPage=1&resultsPerPage=60 |
| ThermoFisher | https://www.thermofisher.com/de/de/home/life-science/protein-biology/protein-labeling-crosslinking/protein-crosslinking/crosslinker-selection-guide.html |
| ThermoFisher | https://www.thermofisher.com/de/de/home/life-science/protein-biology/protein-labeling-crosslinking/protein-crosslinking.html |
| ThermoFisher | https://www.thermofisher.com/de/de/home/life-science/protein-biology/protein-biology-learning-center/protein-biology-resource-library/pierce-protein-methods/chemistry-crosslinking.html |
| ThermoFisher | https://www.thermofisher.com/de/de/home/life-science/protein-biology/protein-labeling-crosslinking/protein-crosslinking/crosslinker-application-reference-guide.html |
| ThermoFisher | https://www.thermofisher.com/de/de/home/life-science/protein-biology/protein-labeling-crosslinking/protein-labeling/biotinylation/photoreactive-biotinylation-reagents.html |
| ThermoFisher | https://www.thermofisher.com/de/de/home/life-science/protein-biology/protein-labeling-crosslinking/protein-crosslinking/crosslinker-selection-tool.html |
| ThermoFisher | https://www.fishersci.com/us/en/products/I9C8L0FC/crosslinking-labeling-protein-modification.html |
| CovaChem | http://www.covachem.com/protein_crosslinkers.html |
| CreativeMolecules | http://www.creativemolecules.com/CM_Products.htm |
| ProteoChem | http://www.proteochem.com/proteincrosslinkers-c-1.html |
| G Biosciences | http://www.gbiosciences.com/ResearchProducts/Protein-Cross-linking.aspx |
| G Biosciences | http://www.genotech.com/protein-research/protein-crosslinking-modification.html |
| G Biosciences | http://www.gbiosciences.com/Protein-Research/Cross-Linking-Modification/Protein-Cross-Linkers |
| Chemometec | http://shop.chemometec.com/?s=cross-linker&post_type=product |
| Interchim | http://www.interchim.eu |
| Biocompare | http://www.biocompare.com/Protein-Biochemistry/11119-Chemicals/?search=crosslinker |
| BioVision | http://www.biovision.com/search/results.html?keywords=crosslinker&x=0&y=0 |
| Abcam | http://www.abcam.com/products?keywords=crosslinker |
| SoltecVentures | http://www.soltecventures.com/PhotoreactivePhenylAzides-c-8 |
| MolBio | http://www.molbio.com/Heterobi.htm |
| MolBio | http://www.molbio.com/HomoNHS.htm |

| | | |
|---|---|---|
| VWR | https://de.vwr.com/store/product/18498564/cross-linkers-heterobifunctional | |
| VWR | http://www.molbio.com/Homomaleim.htm | |
| VWR | http://www.molbio.com/photoact.htm | |
| GooglePatents | https://www.google.de/?tbm=pts&gws_rd=cr&ei=LCjIV6a7CYnWsAG2vIYw#tbm=pts&q=cross-linker | |
| GooglePatents | https://www.google.de/?tbm=pts&gws_rd=cr&ei=LCjIV6a7CYnWsAG2vIYw#tbm=pts&q=crosslinker | |
| ChemicalBook | http://www.chemicalbook.com/Search_EN.aspx?keyword=crosslinker | |
| ChemicalBook | http://www.chemicalbook.com/ChemicalProductProperty_EN_CB9142806.htm | |
| ChemicalBook | http://www.chemicalbook.com/Search_EN.aspx | |
| ChemCD | http://www.chemcd.com/search?q=cross-linker | |
| BioWorld | https://www.bio-world.com/index.php?main_page=advanced_search_result&search_in_description=7&keyword=cross-linker&x=29&y=16 | |
| VWR | https://de.vwr.com/store/product/18498564/cross-linkers-heterobifunctional | |
| CaymanChem | https://www.caymanchem.com/Search?q=crosslinker | |
| SoluLink | http://www.solulink.com/search/google?cx=001277416618431018690%3Ahgffxulp8iy&cof=FORID%3A11%3BNB%3A1&query=crosslinker&op=Go&form_build_id=form-7d49ffbea69c5b09c17b3c3c176a8bd0&form_token=fd2e01ee5d8c6941c678b5571f6972df&form_id=google_cse_searchbox_form | |
| TRC (Toronto Research Chemicals) | http://www.trc-canada.com/products-listing/ | |

Supplementary Table S2: Used vendor publications.

| Vendor | Publication, available on the web as PDF files | web page (accessed 02/2020) |
|---|---|---|
| Creative Molecules Inc. | Crosslinking reagents test reaction procedure | http://www.creativemolecules.com |
| Creative Molecules Inc. | Isotopically-coded crosslinking reagents from Creative Molecules Inc., 2009 | http://www.creativemolecules.com |
| Creative Molecules Inc. | Crosslinking reagents masses | http://www.creativemolecules.com |
| GBiosciences | Protein Cross Linker Selection Guide | http://www.gbiosciences.com |
| GBiosciences | Protein Cross-Linkers Handbook & Selection Guide | http://www.gbiosciences.com |
| GBiosciences | Protein Cross-Linking Techer's Guidebook (Cat. # BE-605) | http://www.gbiosciences.com |
| HiChrom | Gas Chromatography GC Derivatization Reagents | http://www.hichrom.com |
| Invitrogen | Molecular Probes Handbook, Chapter 5: Crosslinking and Photoactivatable Reagents, 11$^{th}$. ed., 2010 | https://www.thermofisher.com |
| Interchim | Derivatization reagents | http://www.interchim.com |
| Mobitec | Chapter 5: Crosslinking and Photoreactive Reagents | http://www.mobitec.com/cms/index.html |
| Pierce | Cross-Linking Reagents, Technical Handbook, 2009 | https://www.thermofisher.com |
| Regis | GC Derivatization Reagents | http://www.registech.com |
| Sigma Aldrich | Derivatization reagents: For selective response and Detection in Complex Matrices | https://www.sigmaaldrich.com |
| Sigma Aldrich | Guide to Derivatization Reagents for GC | https://www.sigmaaldrich.com |
| Thermo Scientific | Avidin-Biotin Technical Handbook | https://www.thermofisher.com |
| Thermo Scientific | Crosslinking Technical Handbook, 2012 | https://www.thermofisher.com |
| Thermo Scientific | Thermo Scientific Pierce Reagents, Solvents and Accessories | https://www.cromlab.es |

Supplementary Table S3: Used vendor-specific cross-linker selection tools resp. guides.

| Vendor | Link to cross-linker selection tool / guide (accessed 02/2020) |
|---|---|
| Thermo Scientific | https://www.thermofisher.com/de/de/home/life-science/protein-biology/protein-labeling-crosslinking/protein-crosslinking/crosslinker-selection-tool.html |
| Thermo Scientific | https://www.thermofisher.com/de/de/home/life-science/protein-biology/protein-labeling-crosslinking/protein-crosslinking/crosslinker-selection-guide.html |
| GBiosciences | http://info.gbiosciences.com/blog/bid/160707/What-is-Protein-Cross-Linking-and-Which-Reagents-are-Used-in-it |

Supplementary Table S4: Amino acid functional groups amenable to reactive groups of cross-linkers

| Amino acid | Functional group of amino acid … | … is reacting primarily with the following reactive groups of cross-linkers |
|---|---|---|
| S - serine | hydroxyl (-OH) | isocyanate, vinyl sulfone |
| T - threonine | | |
| Y - tyrosine | | |
| K - lysine | amine (–NH2), amides | (form)aldehyde, aryl halide, carbodiimide, carboxyl, hydroxybenzotriazole, phosphine, imidoester, N-hydroxyphthalimide, NHS ester, Sulfo-NHS ester, pentafluorophenyl, thioimidate, vinyl sulfone |
| N - asparagine | | |
| Q - glutamine | | |
| R - arginine | guanidino | glyoxal, phenylglyoxal, formaldehyde |
| C - cysteine | sulfhydryl -SH | acryl, haloacetyl (bromo-, iodo-), platinum(II), maleimide, methanethiosulfonate, pyridinyldisulfide, vinyl sulfone |
| D – aspartic acid | carboxyl (-COOH) | amine, carbodiimide, dimethylether |
| E – glutamic acid | | |
| M - methionine | thioether (–S-) | platinum(II) |
| H - histidine | imidazole | platinum(II), formaldehyde |
| W - tryptophan | indole | formaldehyde |
| F - phenylalanine | phenyl | formaldehyde |

Supplementary Table S5: Examples of typical spacer arm components, which are often used in cross-linking reagents

| Spacer arms |
|---|
| 6-amino-caprionic acid |
| DADPA (diaminodipropylamine) |
| ethylenediamine |
| hexanediamine |
| $(PEG)_n$ - polyethylenglycol |

Supplementary Table S6: Handles for detection, affinity enrichment or purification used in cross-linking reagents

| Handle | Detection resp. enrichment / purification with … |
|---|---|
| Biotin | avidin, streptavidin or neutravidin |
| alkyne | alkyne-azido click chemistry or immobilized azide linkers |
| azide (-N3) | immobilized alkyne linkers or click chemistry using biotin-alkyne respective biotin-phosphine |
| CHCA (cyano-4-hydroxycinnamic acid) | UV-absorbing label for signal enhancement |
| DNB (dinitrobenzole) | anti-DNB antibodies |

Supplementary Table S7: Term statistics of the XLMOD.obo ontology

| Term type | Number of terms |
|---|---|
| Typedef's | 16 |
| Cross-linking attributes | 16 |
| Handles | 10 |
| Reactive group terms | 39 |
| Reactive group attributes (without reactivities) | 8 |
| Reactivity terms | 37 |
| Cross-linker related chemical modification terms | 103 |
| Zero-length cross-linker | 6 |
| Bifunctional cross-linker terms | 248 |
| Trifunctional cross-linker terms | 13 |
| Label transfer reagents | 34 |
| Amino acid derivatives | 5 |
| Amino acids | 4 |
| GC/MS derivatization reagents | 174 |
| LC/MS derivatization reagents | 324 |
| other terms | 13 |
| Nucleobases and their derivatives | |
| **Total number of terms** | **1083** |

Supplementary Table S8: ID ranges for the terms in the XLMOD.obo ontology

| Term type | ID range starting with |
|---|---|
| General terms, reactivities, handles, other properties | xlmod:00000 |
| Concrete reactive groups and their attributes | xlmod:00100 |
| Mono-functional cross-linker (cross-linker related chemical modifications) | xlmod:01000 |
| Label transfer reagents | xlmod:01800 |
| Analogues of amino acids | xlmod:01900 |
| Bi-functional cross-linkers | xlmod:02000 |
| Tri-functional cross-linkers | xlmod:03000 |
| Derivatization reagents | xlmod:06000 |
| Further reactivities (used for GC/MS and LC/MS reagents) | xlmod:06500 |
| GC/MS derivatization reagents | xlmod:07000 |
| LC/MS derivatization reagents | xlmod:09000 |